\pdfoutput=1
\documentclass{ws-rv975x65}
\usepackage{anonchap}
\usepackage{subfigure}     
\usepackage[square]{ws-rv-van}     
\makeindex
\renewcommand*\thesection{\arabic{section}}
\renewcommand*\thefigure{\arabic{figure}}

\begin{document}

\chapter*{The Parton Model and its Applications}
\emph{(Contribution to a book to be published by World Scientific for the occasion of 50 Years of Quarks)}\\ \\

\author[Tung-Mow Yan \& Sidney D. Drell]{Tung-Mow Yan\footnote{Laboratory for Elementary Particle Physics, Cornell University, Ithaca, NY 14853} and Sidney D. Drell\footnote{SLAC National Accelerator Laboratory, Stanford University, Stanford, CA 94305}}


\begin{abstract}
This is a review of the program we started in 1968 to understand and generalize Bjorken scaling and Feynman's parton model in a canonical quantum field theory. It is shown that the parton model proposed for deep inelastic electron scatterings can be derived if a transverse momentum cutoff is imposed on all particles in the theory so that the impulse approximation holds. The deep inelastic electron-positron annihilation into a nucleon plus anything else is related by the crossing symmetry of quantum field theory to the deep inelastic electron-nucleon scattering. We have investigated the implication of crossing symmetry and found  that the structure functions satisfy a scaling behavior analogous to the Bjorken limit for deep inelastic electron scattering. We then find that massive lepton pair production in collisions of two high energy hadrons can be treated by the parton model with an interesting scaling behavior for the differential cross sections. This turns out to be the first example of a class of hard processes involving two initial hadrons. 
\end{abstract}
\body

\section{Introduction}
In the 1950s and 1960s many new particles were found experimentally. M. Gell-Mann \cite{gell61} and Y. Ne’eman \cite{neem61} showed that these particles can be fit into an octet (mesons and baryons) or a decuplet (hyperons) representation of a symmetry group called SU(3). A peculiar feature is that the simplest representation of the group, a triplet, was not realized in Nature. Furthermore, it is hard to imagine that these particles were all elementary. Gell-Mann \cite{gell64}  and Zweig \cite{zwei64} independently discovered that if one proposes the existence of three spin 1/2 fundamental particles which Gell-Mann called quarks, then a meson can be treated as a bound state of a quark and an antiquark, and a baryon can be treated as a bound state of three quarks. To accomplish this feat, however, the fundamental constituents quarks must possess very strange properties: their electric charge must be fractional (1/3 or 2/3 of an electronic charge), and they must violate the spin-statistics connection. Later, the new quantum number color \cite{gree64} was proposed to resolve these difficulties. This was the birth of the quark model in 1964. 

In the mean time, the popularity of canonical quantum field theories was in decline due to the absence of a viable field theory for the strong interactions. Instead, Gell-Mann \cite{gell64b} postulated that the commutation relations derived from spin 1/2 quark fields for the vector and axial vector currents are exact whether or not the quarks exist and whether the underlying symmetry $\text{SU}(3) \times \text{SU}(3)$ is exact or not. These current algebras, in combination with PCAC (partially conserved axial current) \cite{namb60} and soft pion theorems provide a framework for extracting dynamical information on strong and weak interactions. Initial applications focused on low energy phenomena. The first application of current algebra to high energy processes was made by S. Adler \cite{adle66} who derived sum rules for high energy neutrino and anti-neutrino scatterings. At the time the prospect for neutrino scatterings was quite remote. J. Bjorken \cite{bjor66} obtained from these sum rules  an inequality for high energy electron nucleon scatterings by an isospin rotation. The inequality showed that the cross sections for electron nucleon scatterings is of comparable size with that for a point-like target, and this could be tested by the ongoing SLAC-MIT experiment at SLAC. 

In an attempt to understand Adler's sum rules and Bjorken's inequality, Bjorken proposed that the structure functions that describe the cross-sections for the inelastic electron scatterings satisfy a scaling property known as Bjorken scaling \cite{bjor69}. There are two Lorentz invariant kinematic variables for the inelastic electron scatterings. The structure functions depend on the two variables. Bjorken scaling means that in the large momentum transfers, these structure functions become a function of the ratio of the two variables. Bjorken scaling was quickly confirmed by the experiments at SLAC \cite{pano68}. Feynman \cite{feyn69} interpreted the Bjorken scaling as the point-like nature of the nucleon's constituents when they were incoherently scattered by the incident electron. Feynman named the point-like constituents partons. This is the parton model. Feyman left open the possibility that the partons need not be the quarks. However, theorists quickly identified the partons with quarks (in the late 1960s and early 1970s QCD did not exist, and so gluons did not enter the picture). A nucleon consists of three ``valence'' quarks which carry the nucleon's quantum numbers and a ``sea'' of quark-antiquark pairs. This identification led to many predictions for electron and neutrino (and antineutrino) scatterings from a nucleon \cite{clos79}. 

In the fall of 1968 soon after Feynman’s parton model was proposed, Don Levy and we embarked on a comprehensive program \cite{drel69, drel69b, drel70, drel70b, drel70c} to understand and apply the parton model in a quantum field theory framework. First, we would like to know under what conditions the parton model could be derived from a quantum field theory. Second, was it possible to apply the parton model to other processes? Let us state briefly the conclusions of our investigations here. We showed that the parton model could be derived if the impulse approximation was valid: that during the scattering the constituents behave as if they were free. To accomplish this, we had to impose a transverse momentum cut-off for the particles that appeared in the quantum field theory \cite{drel69, drel69b, drel70}. Crossing symmetry in quantum field theory relates deep inelastic electron nucleon scatterings and deep inelastic  electron-positron annihilation into a nucleon plus anything else. We have found  the parton model can be applied to  the crossed channel reaction. The structure functions are found to have Bjorken scaling  as in the scattering case. In search for other processes to apply the parton model, we found at least one: namely, the lepton pair production by proton-proton collision which was under study by Christenson, et al \cite{chri70} at BNL. The conditions for applying the parton model were satisfied if the lepton pair was produced by the  annihilation of  a parton from one proton and an anti-parton from the other proton and the lepton pair mass is sufficiently high. This is now known as Drell-Yan mechanism \cite{drel70d}.

Finally, we must specify a particular quantum field theory. At the time, there was no good candidate for a quantum field theory for the strong interactions. So our choice to a large extent is arbitrary. We were guided by simplicity. If we were to include particles with isospin, then we had to exclude vector mesons. Otherwise, we would have to deal with  non-Abelian massive vector mesons, and no one knew how to do that. Thus, we settled on a quantum field theory involving a nucleon isodoublet and an isotriplet pions, so that there will be both a  spin 1/2 charge carrier and a spin 0 charge carrier. They interact through a pseudoscalar coupling. 

In the following we will apply our model quantum field theory to elaborate on  the points we mentioned above. It should be pointed out that our emphasis will be on the motivations,  principles,  the most general results, and what modifications that have to be made after the discovery of QCD. We should also mention that we have also  found a relation between the threshold behavior of the structure functions in the deep inelastic scatterings and the asymptotic behavior of the elastic electromagnetic form factor \cite{drel70e}, and have  studied the semi-inclusive deep inelastic scatterings $e + P \rightarrow  e' + h +X$ which gives rise to fragmentation functions \cite{drel70f}. But we will not discuss these last two topics any further. 

\section{Derivation of the Parton Model From a Canonical Quantum Field Theory}\label{sec:partonmodel}
\begin{figure}
\centerline{\includegraphics[width=0.6\textwidth]{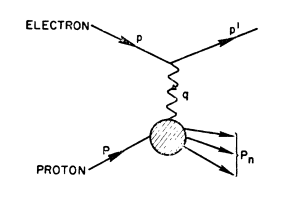}}
\caption{}\label{fig:epkin}
\end{figure}

The differential cross-section for the inelastic electron scattering from a nucleon
\begin{equation}\label{equ:inelasticscat}
e + P \rightarrow e' + \text{ anything}
\end{equation}
is described by the tensor:
\begin{align}\label{equ:Wtensor}
W_{\mu \nu} &= 4 \pi^2 \frac{E_p}{M} \sum_n \left< P | J_\mu (0) |n \right> \left< n | J_\nu (0) | P \right> (2 \pi)^4 \delta^4 (q + P - P_n)\notag\\
&= - \left( g_{\mu \nu} - \frac{q_\mu q_\nu}{q^2} \right) W_1 (q^2, \nu)\\
&~~~~~~~~ + \frac{1}{M^2} \left( P_\mu - \frac{P \cdot q}{q^2} q_\mu \right)  \left( P_\nu - \frac{P \cdot q}{q^2} q_\nu \right)  W_2(q^2,\nu)\notag
\end{align}
where $\left|P\right>$ is a one nucleon state with four momentum $P$, $J$ is the electromagnetic current, $q$ is the four momentum of the virtual photon, $q^2 = - Q^2  < 0$  is the square of the virtual photon mass, and 
$M\nu = P\cdot q$ is the energy transfer to the photon in the laboratory system. An average over the proton spin is understood.  The kinematics is depicted in Fig. \ref{fig:epkin}. 

The differential cross section in the rest frame of the proton is given by 
\begin{equation}
\frac{d^2 \sigma}{d \epsilon' d \cos \theta} = \frac{8 \pi \alpha^2}{(Q^2)^2} (\epsilon ')^2 \left[ W_2(q^2, \nu) \cos^2(\theta/2) + 2 W_1(q^2,\nu)\sin^2(\theta/2) \right]
\end{equation}
where $\epsilon$ and $\epsilon'$ are the initial and final energy of the electron and $\theta$ is its scattering  angle.

We are interested in the Bjorken limit of large $Q^2$ and $\nu$ with the ratio $\xi= Q^2/2M\nu$ fixed. Let us work in the infinite-momentum center of mass frame of the electron and proton. Then
\begin{equation}
q^0 = \frac{2M \nu - Q^2}{4 P}, ~~~~ q_3 = \frac{-2 M \nu - Q^2}{4 P}, ~~~~ |\mathbf{q_\bot}| = \sqrt{Q^2} + \mathcal{O}(1/P^2),
\end{equation}
with the nucleon momentum $\mathbf{P}$ along the 3-axis. We go to the interaction picture with the familiar U-matrix transformation
\begin{equation}
J_\mu(x) = U^{-1}(t) j_\mu (x) U(t),
\end{equation}
where $J(x)$ and $j(x)$ are the fully interacting and bare electromagnetic currents, respectively. 

\begin{figure}
\centerline{\includegraphics[width=0.8\textwidth]{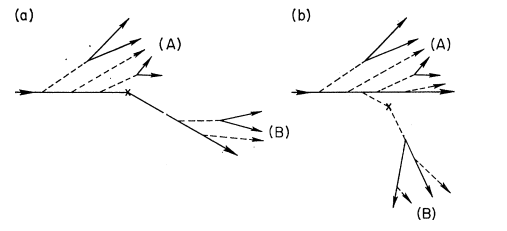}}
\caption{}\label{fig:partsplit}
\end{figure}

Eq. (\ref{equ:Wtensor}) can be rewritten as 
\begin{align}
W_{\mu \nu} =\, &4 \pi^2 \frac{E_p}{M} \sum_n \left< U P \right| j_\mu (0)U(0) \left|n\right>\left<n\right|U^{-1}(0)j_\nu(0)\left|UP\right>\notag\\
&\times (2 \pi)^4 \delta^4(q + P - P_n),\label{equ:Wtensor2}
\end{align}
where $\left|U P\right> = U(0)\left|P\right>$. In the old fashioned perturbation theory, the state $\left|U P\right>$ is represented by all the particles that appear just before the electromagnetic vertex, and the state $U(0)\left|n\right>$ is represented by all the particles that appear right after the electromagnetic vertex (see Fig. \ref{fig:partsplit}).

From Fig. \ref{fig:partsplit}, it is seen that there are two groups of particles after the scattering by the photon, group (A) moves along the direction of the initial nucleon momentum $\mathbf{P}$, and group (B) moves along the scattered momentum $\mathbf{p} + \mathbf{q}$, where $\mathbf{p}$ is the momentum of the parton to be scattered by the photon. Each group will have limited transverse momentum relative to their large momentum $\mathbf{P}$ and $\mathbf{p} + \mathbf{q}$, respectively.
 
If we schematically denote the energy of the a particular component of $\left|UP\right>$  by $E_{UP}$, and the energy of a particular component of $U(0)\left|n\right>$ by $E_{Un}$, then with a transverse momentum cutoff introduced  for each field particle in the theory, the energy differences $E_P  - E_{UP}$ and $E_n - E_{Un}$ are
\begin{subequations}
\begin{eqnarray}
E_P - E_{UP} &= \mathcal{O}\left(\frac{k_T^2 + M^2}{P}\right)\label{equ:deltaE1},\\
E_n - E_{U n} &= \mathcal{O}\left(\frac{k_T^2 + M^2}{P}\right)\label{equ:deltaE2},
\end{eqnarray}
\end{subequations}
where $k_T$ and $M$ are typical transverse momentum and mass scales, and Eqs. (\ref{equ:deltaE1}), (\ref{equ:deltaE2}) are small compared with $q^0$, so we can make the substitution in the overall energy momentum conserving Dirac delta function in Eq. \ref{equ:Wtensor2}, 
\begin{equation}\label{equ:energy}
q^0 + E_P - E_n = q^0 + E_{U P} - E_{U n} = q^0 + E_a - E_{a'},
\end{equation}
where $a$ is a constituent in $\left|UP\right>$ with momentum $p_a$  which is scattered into a constituent $a'$ with momentum $p_{a'}$. In other words, the overall energy conservation implies that the energy is conserved across the electromagnetic vertex. We should emphasize that we have accomplished this by imposing a transverse momentum cutoff in the underlying quantum field theory and the Bjorken limit gives $Q^2$, and $M\nu \gg k_T^2$. Eq. (\ref{equ:energy}) is a statement of impulse approximation, because the scattering event occurs so suddenly that the constituents involved can be treated as free. We now make use of the translation operators, completeness of the states $\left|n\right>$ and unitarity of the $U$ matrix to obtain:
\begin{align}
\lim_{P\rightarrow \infty; q^2, M\nu \rightarrow \infty, \omega \text{ fixed}}& W_{\mu \nu}\notag\\
=\,&4 \pi^2 \frac{E_p}{M}\sum_n\int (dx) e^{+i q x}\left<UP\right|j_\mu (x) U(0) \left|n\right>\notag\\
&\times\left<n\right|U^{-1}(0)j_\nu(0)\left|UP\right>\notag\\
=\,&4 \pi^2 \frac{E_p}{M}\int (dx) e^{+i q x}\left<UP\right|j_\mu (x) U(0) U^{-1}(0) j_\nu (0) \left|UP\right>\notag\\
=\,&4 \pi^2 \frac{E_p}{M}\int (dx) e^{+i q x}\left<UP\right|j_\mu (x) j_\nu (0) \left|UP\right>.\label{equ:limW}
\end{align}

Energy-momentum conservation across the electromagnetic vertex gives                                                                                                                                      
\begin{equation}\label{equ:enmom}
(p_a + q)^2  =(p_{a'})^2  = 0,
\end{equation}
or
\begin{equation*}
2 p_a \cdot q + q^2 = 0.
\end{equation*}
If we denote by $\xi$ the fraction of the longitudinal momentum carried by the initial parton,
\begin{equation}
p_a = \xi  P,     
\end{equation}
then
\begin{equation}\label{equ:xidef}
\xi = \frac{1}{\omega} = \frac{Q^2}{2M\nu}.
\end{equation}

Thus, the Bjorken scaling variable is identified as the fractional longitudinal momentum carried by the scattered parton. The structure functions $W_1$ and $W_2$ are related to the longitudinal momentum distribution functions of the initial proton. Working out the tensor structure of  Eq. (\ref{equ:limW}), we find that the structure functions  $W_1$ and $W_2$ depend only on the Bjorken variable $\omega$,
\begin{subequations}
\begin{eqnarray}
M W_1 (q^2, \nu) &= F_1(\omega),\\
W_2(q^2, \nu) &= F_2(\omega),
\end{eqnarray}
\end{subequations}
and the relations between $W_1$ and $W_2$ depending on the spin of the current, 
\begin{subequations}
\begin{align}\label{equ:DISFF}
F_1(\omega) &= \frac{1}{2} \omega F_2(\omega), &\text{spin 1/2 current},\\
F_1(\omega) &= 0, &\text{spin 0 current}.
\end{align}
\end{subequations}
This completes the derivation of the parton model. 

\section{QCD and the Improved Parton Model}
Our studies preceded the discovery of Quantum Chromodynamics (QCD) which is a non-Abelain gauge theory \cite{yang54} with octet colored gluons and triplet color quarks (at present, there are 3 generations of triplet colored quarks: $u$, $d$, $c$, $s$, $t$, and $b$). It has the unique property of asymptotic freedom \cite{gros73} that its coupling constant decreases logarithmically with momentum scale $Q$. Since 1973, QCD has been accepted as the correct theory of strong interactions. In this theory deep inelastic scatterings can be analyzed rigorously. The main results are:

\begin{arabiclist}[]
\item{The moments of the structure functions $W_1$, and $W_2$ are no longer independent of $Q$ as they would be if Bjorken scaling is exact. These moments decrease with $Q$ logarithmically with certain powers which are called anomalous dimensions of twist 2 operators and are calculable in QCD \cite{chen84}.}
\item{It is possible to relate the more formal analysis of QCD to the more intuitive parton model. The relations are provided by a set of Altarelli-Parisi equations \cite{alta77}. These equations describe, as we increase $Q$, how a gluon evolves into a gluon pair or a quark-antiquark pair, or a quark evolves into a quark plus a gluon. These evolutions are described by a set of quantities which are called splitting functions. Moments of these splitting functions turn out to be the anomalous dimensions which appear in the moments of the structure functions $W_1$ and $W_2$. Thus, the Altarelli-Parisi equations offer further insight into the parton model and the working of QCD. }
\end{arabiclist}

\section{Deep Inelastic Electron-Positron Annihilation}
\begin{figure}
\centerline{\includegraphics[width=0.5\textwidth]{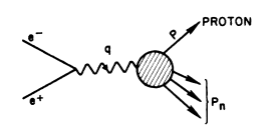}}
\caption{}\label{fig:eekin}
\end{figure}

The process
\begin{equation}\label{equ:eeannih}
e^+ + e^- \rightarrow P + \text{ anything}
\end{equation}
is related to the inelastic electron scattering (\ref{equ:inelasticscat}) by the crossing symmetry of relativistic quantum field theory. The kinematics is shown in Fig. \ref{fig:eekin}.

It is therefore interesting to ask if the parton model ideas can be applied to this process. We will indicate in this section that the answer indeed is yes. We will only sketch the main ideas. The details can be found in \cite{drel69b, drel70b}. The cross section for the process  (\ref{equ:eeannih}) is summarized by the two structure functions defined by \cite{drel69b, drel70b}
\begin{align}
\bar{W}_{\mu \nu} =\, &4 \pi^2 \frac{E_p}{M} \sum_n \left< 0 \right| J_\mu (0) \left|P n\right>\left<nP\right|J_\nu(0)\left|0\right> \notag\\
&\times(2 \pi)^4 \delta^4(q-P-P_n)\notag\\
=&-\left(g_{\mu\nu} - \frac{q_\mu q_\nu}{q^2} \right) \bar{W}_1(q^2,\nu)\notag\\
&+ \frac{1}{M^2}\left(P_\mu -\frac{P \cdot q}{q^2}q_\mu \right)\left(P_\nu - \frac{P \cdot q}{q^2} q_\nu\right)\bar{W}_2(q^2,\nu).
\end{align}
Then the differential cross section is given by \cite{drel69b, drel70b}
\begin{multline}
\frac{d^2 \sigma}{dE d\cos \theta}= \frac{4 \pi \alpha^2}{(q^2)^2} \frac{M^2 \nu}{\sqrt{q^2}}\left(1-\frac{q^2}{\nu^2}\right)^{1/2} \bigg[ 2 \bar{W}_1(q^2,\nu)\\ + \frac{2 M \nu}{q^2} \left(1-\frac{q^2}{\nu^2}\right) \frac{\nu \bar{W}_2(q^2,\nu)}{2 M} \sin^2\theta \bigg]
\end{multline}
where $E$ is the energy of the detected proton and $\theta$ is the angle of the proton momentum $\mathbf{P}$ with respect to the colliding $e^+$ and $e^-$ beams in the center of mass system. The two Lorentz invariant  kinematic variables are defined by 
\begin{subequations}
\begin{align}
   Q^2 &=q^2  > 0,\\
  M \nu &= P\cdot q . 
 \end{align}
 \end{subequations}
In the present case the ratio $0< 2M\nu/Q^2<1$. If we follow a similar analysis given to deep inelastic scattering in Section \ref{sec:partonmodel}, we will find that the two structure functions satisfy Bjorken scaling: 
\begin{align}
\lim\nolimits_{\text{Bj}} M \bar{W}_1 (q^2, \nu) &= -\bar{F}_1(\omega),\notag\\
\lim\nolimits_{\text{Bj}} \nu \bar{W}_2 (q^2, \nu) &= \bar{F}_2(\omega),
\end{align}
where $\omega = 2 M\nu /q^2$. Furthermore, we have
\begin{subequations}
\begin{align}
\bar{F}_1(\omega)/\bar{F}_2(\omega) &= \frac{1}{2} \omega &\text{(spin-1/2 current)}\\
\bar{F}_1(\omega) &= 0 &\text{(spin-0 current)}
\end{align}
\end{subequations}

The above relations are similar to the results Eq. (\ref{equ:DISFF}) for deep inelastic scattering. In the deep inelastic scattering case, the Bjorken's scaling variable $\xi$ is identified with the longitudinal momentum fraction carried by the scattered parton inside the nucleon. What is the meaning for the corresponding scaling variable $Q^2/2 M \nu$ in the deep inelastic annihilation? Let us follow similar steps as in Eqs. (\ref{equ:enmom})-(\ref{equ:xidef}). The annihilation process proceeds through the creation of a parton $a$ and its anti-parton $a'$, followed by the decay of parton $a$ into a group of particles which contains the detected nucleon with momentum $P$. Energy-momentum conservation at the electromagnetic vertex gives,
\begin{equation}
(q - p_a )^2  = p_{a'}^2,
\end{equation}
or
\begin{equation}\label{equ:enmomann}
Q^2 -2p_a\cdot q = 0.
\end{equation}

If we denote by $\eta$ the ratio of the momentum $p_a$ of the parent parton $a$ to the nucleon momentum $P$,
\begin{equation}
p_a = \eta P,
\end{equation}
then Eq. (\ref{equ:enmomann}) gives
\begin{equation}
\eta = Q^2 / 2M \nu > 1.
\end{equation}
The ratio is larger than unity as it should be since the nucleon can only carry a fraction of the momentum of its parent parton $a$. Finally, these scaling predictions will receive QCD's logarithmic corrections in $Q$ \cite{muel78}, just as in the case of deep inelastic scatterings. 

\section{Lepton Pair Production}
The field on lepton pair production began with the experiment at BNL by Christenson et. al. \cite{chri70}. They studied the reaction
\begin{equation}\label{equ:pairprod}
p + U \to \mu^+ \mu^- + X
\end{equation}
for proton energies 22-29 GeV, and the muon pair mass 1-6.7 GeV. Two features of the data stand out: (1) the shoulder-like structure near the muon pair mass of 3 GeV, and (2) the rapid fall-off of the cross section with the muon pair mass. We know now that the shoulder-like structure is due to the  $J/\Psi$ which was discovered in 1974 by a muon pair production experiment at BNL \cite{albe74}  and an $e^+ e^-$ colliding beam experiment at SLAC \cite{augu74}.

We got interested in the process (\ref{equ:pairprod}) for two reasons: (1) we were looking for applications of the parton model outside deep inelastic lepton scatterings, and (2) we wanted to understand if the rapid decrease of the cross section with the muon pair mass could be reconciled with the point-like cross sections observed in the deep inelastic electron scatterings. 

\begin{figure}
\centerline{\includegraphics[width=0.8\textwidth]{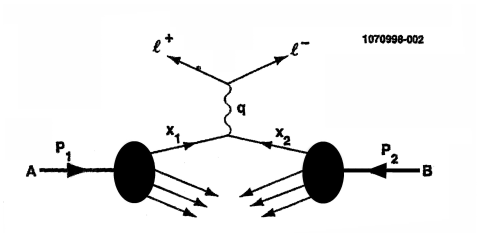}}
\caption{}\label{fig:llprod}
\end{figure}

The key idea in our approach was once again the impulse approximation. First, we picked an appropriate infinite momentum frame to exploit the time dilation. In this frame, if we were able to establish that the time duration of the external probe is much shorter than the lifetimes of the relevant intermediate states, i.e.
\begin{equation}\label{equ:taufact}
\tau_\text{probe} \ll \tau_\text{int. states},
\end{equation}
then the constituents could be treated as free. Thus, the cross section in the impulse approximation is a product of the probability to find the particular parton configuration and the cross section for the free partons. In the case of lepton pair production from two initial hadrons,
\begin{equation}
P_1 + P_2 \to l^+ l^- + X,
\end{equation}
the pair production by the parton-antiparton annihilation satisfies the criteria of impulse approximation \cite{drel70d} (see Fig. \ref{fig:llprod}).

It is easily shown that the fractional longitudinal momenta of the annihilating partons satisfy  
\begin{equation}
\tau = x_1 x_2 = \frac{Q^2}{s},
\end{equation}
where $Q^2$ and $s$ are, respectively, the pair mass squared and the square of the C.M. energy of the initial energy of the initial hadrons. The rapidity of the pair is given by   
\begin{equation}
y=\frac{1}{2}\ln\frac{x_1}{x_2}.
\end{equation}

The predictions stated in our original paper \cite{drel70d} are         
\begin{arabiclist}[]
\item{The magnitude and shape of the cross section are determined by the parton and antiparton distributions measured in deep inelastic lepton scatterings:
\begin{equation}
\frac{d \sigma}{dQ^2 dy} = \frac{4 \pi \alpha^2}{3 Q^4} \frac{1}{N_c} \sum_p x_1 fp(x_1) x_2 f \bar{p}(x_2)
\end{equation}
where a color factor $N_c$ is included in anticipating QCD;}
\item{The cross section $Q^4 d\sigma/dQ^2$ depends only on the scaling variable $\tau =Q^2/s$;}          
\item{If a photon, pion, kaon, or antiproton is used as the projectile, its structure functions can be measured by lepton pair production \cite{chan13}. This is the only way we know of to study the parton structure of a particle unavailable as a target for lepton scatterings;}
\item{The transverse momentum of the pair should be small ($\sim$ 300-500 MeV);}
\item{In the rest frame of the lepton pair, the angular distribution is $1 + \cos^2\theta$ with respect to the hadron collision axis, typical of the spin 1/2 pair production from a transversely polarized virtual photon;}
\item{The same model can be easily modified to account for $W$ boson productions.}
\end{arabiclist}
In this model, the rapid decrease of the cross section with $Q^2$ as seen in (\ref{equ:pairprod}) is related to the rapid fall-off of structure functions as $x \rightarrow 1$ in deep inelastic electron scatterings. 

The lepton pair production considered here is the first example of a class of hard processes involving two initial hadrons. These processes are not dominated by short distances or light cone. So the standard analysis using operator product expansion is not applicable. But the parton model works. Soon after our work, Berman, Bjorken, and Kogut \cite{berm71} applied similar ideas to large transverse momentum processes
\begin{equation}
h_1 + h_2 \to h(\text{large } P_T) + X
\end{equation}
induced by deep inelastic electromagnetic interactions. At that time, it was believed that strong interactions severely suppressed large transverse momenta, therefore, electromagnetic interactions would quickly dominate the large transverse momentum processes. This was the precursor of the point-like gluon exchanges in QCD.

After the advent of QCD, the basic picture of lepton pair production has been confirmed theoretically and the details have been greatly improved \cite{ster95}. It is no longer a model. That lepton pairs are produced by parton-antiparton annihilation is a consequence of QCD. In QCD, the partons are quarks, antiquarks, and gluons, and the number of color $N_c =3$. The unique property of QCD being an asymptotically free gauge theory makes the parton model almost correct, namely for deep inelastic processes we have
\begin{equation}
\text{QCD}=\text{parton model + small corrections}.
\end{equation}

In the modern language, the impulse approximation is replaced by the more precise concept of factorization which separates the long distance and short distance physics and the condition (\ref{equ:taufact}) now becomes   
\begin{equation}
Q^2 \gg \Lambda^2_\text{QCD},
\end{equation}
where $\Lambda_\text{QCD}$ is a typical momentum scale in QCD. The constituents are almost free, leading to logarithmic corrections to the structure functions   
\begin{equation}
f_i \implies f_i\left(x,\ln Q^2\right).
\end{equation}
Factorization for the lepton pair production works in QCD, but in a more complicated manner and it has taken the hard work of many people and many years to establish \cite{coll88, coll11}. The main complication arises from the new feature of initial and final state interactions between the hadrons \cite{neub14}. The result is fairly simple to state
\begin{equation}
\frac{d \sigma^{AB}}{dQ^2 dy} = \sum_{a,b} \int_{x_A}^1 d\xi_A \int_{x_B}^1 d\xi_B f_{a/A}(\xi_A,Q^2) f_{b/B}(\xi_B,Q^2) H_{ab},
\end{equation}
where the sum over $a$ and $b$ are over parton species. The parton distribution functions are the same as those in deep inelastic lepton scatterings with the understanding that $Q^2$ is its absolute value. The function $H_{ab}$ is the parton level hard scattering cross section computable in perturbative QCD and is often written as 
\begin{equation}
H_{ab} = \frac{d \hat{\sigma}}{dQ^2 dy}
\end{equation}
Beside the logarithmic scaling violation, a large transverse momentum of the lepton pair can be produced by recoil of quarks or gluons. A simple dimensional analysis gives    
\begin{equation}
\left< k_T^2 \right> = a + \alpha_s(Q^2) s f(\tau,\alpha_s)
\end{equation}
The constant $a$ is related to the primordial or intrinsic transverse momentum of the partons.

The full angular distributions in both $\theta$ and $\phi$ depend on input quark and gluon densities and are rather complicated \cite{coll77}. For small $k_T$ the $\theta$ dependence is close to $1 + \cos^2 \theta$ even when high order corrections are taken into account. For large $k_T$, the $\theta$ dependence is expected to be substantially modified \cite{peng14}.
 
Many of the predictions have been tested and confirmed by many experiments at Fermilab and CERN and elsewhere \cite{peng14}. We will not go into the details. We will only point out that the model is so successful that its data have become an integral component of the global fit together with the deep inelastic lepton scatterings in determining the parton distributions inside a nucleon. 

\section{The Process as a Tool for New Discoveries}
It seems natural to broaden the definition of Drell-Yan process  to mean a class of high energy hadron-hadron collisions in which there is a subhard process involving one constituent from each of the two incident hadrons. New physics always manifests itself in the production of new particle(s). Since the ordinary particles do not carry the new quantum number of the new physics, to discover new physics in a hadron-hadron collider therefore requires annihilation of the ordinary  particles to create these new particles. Thus, the Drell-Yan mechanism is an  ideal tool for the new discoveries. Let us mention three important discoveries in the recent past which had employed this process to help:

\begin{arabiclist}[]
\item{It was used to design the experiments at CERN that discovered the $W$ and $Z$ bosons \cite{arni83}.}
\item{The process was also crucial in the discovery of the top quark at Fermilab\cite{abe95}.}
\item{The discovery of the Higgs Boson at CERN in 2012 \cite{CMS12} was perhaps the most dramatic example of the utility of the process. The Higgs Boson is the last particle that appears in the Standard Model to have been found. Single Higgs Bosons are predominantly produced by gluon fusion which is a generalized ``Drell-Yan mechanism".}
\end{arabiclist}

Since the first experiment at BNL and the na\"{\i}ve model proposed to understand it, both experiments and theory have come a long way. It is interesting to note that our original crude fit \cite{drel70d} did not remotely resemble the data. We went ahead to publish our paper because of the model's simplicity and our belief that future experiments would be able to definitely confirm or demolish the model. It is gratifying to see that the successor of the na\"{\i}ve model, the QCD improved version, has been confirmed by the experiments carried out in the last forty years. The pair production process has been an important and active theoretical arena to understand various theoretical issues such as infrared divergences, collinear divergences leading to the factorization theorem in QCD for hard processes involving two initial hadrons. The process has been so well understood theoretically that it has become a powerful tool for  discovering new physics. We can expect to find new applications of this process in the future. 

\section{Conclusion}

In this paper we have reviewed the development of our effort to understand Bjorken scaling and Feynman's parton model immediately after these ideas were proposed. We chose to study these topics in a relativistic canonical quantum field theory to take advantage of its crossing symmetry, unitarity, etc. An important input for our program is the impulse approximation. We discovered quickly that the impulse approximation is severely violated due to large logarithms that are present at high energies and large transverse momenta. To restore the validity of the impulse approximation, we had to impose a transverse momentum cutoff for each of the particles in the theory. We then picked the simplest quantum field theory describing nucleons and pions in a pseudoscalar coupling. We started our program in 1968, before QCD and asymptotic freedom were known. Within this framework, we derived the parton model for deep inelastic lepton scatterings, found that the parton model is applicable to crossed channel of  the deep inelastic inclusive electron-positron annihilation into hadrons, and an additional application of the parton model to massive lepton pair production in hadron-hadron collisions. Some of the interesting results are presented in this article. In spite of an unrealistic quantum field theory and the transverse momentum cutoff imposed by hand, most of our results remain valid today except for mild scaling violations in QCD. Our goal was to extract results from general properties of the underlying theory rather than from its specifics. After more than forty years, the subjects that we studied: deep inelastic scatterings, deep inelastic electron-positron annihilation, and lepton pair production, are still very active players in the arena in our quest for  our understanding of the inner structure of the  elementary particles. We were fortunate that we had the opportunity to play a small part in the endeavor. 

\section*{Acknowledgements}
We thank Professor Jen-Chieh Peng and Professor Matthias Neubert for useful communications.

\end{document}